\documentstyle[a4,11pt]{article}

\newcommand{\nit}{\noindent}
\newcommand{\nl}{\newline}
\newcommand{\np}{\newpage}
\newcommand{\dsp}{\displaystyle}
\newcommand{\vs}[1]{\vspace{#1 ex}}
\newcommand{\hs}[1]{\hspace{#1 em}}
\newcommand{\bfr}{\begin{flushright}}
\newcommand{\efr}{\end{flushright}}
\newcommand{\bc}{\begin{center}}
\newcommand{\ec}{\end{center}}
\newcommand{\ben}{\begin{enumerate}}
\newcommand{\een}{\end{enumerate}}

\newcommand{\be}{\begin{equation}}
\newcommand{\ee}{\end{equation}}
\newcommand{\ba}{\begin{array}}
\newcommand{\ea}{\end{array}}
\newcommand{\ct}{\cite}
\newcommand{\bit}{\bibitem}

\newcommand{\gam}{\gamma}

\newcommand{\eps}{\epsilon}

\newcommand{\kg}{\kappa}

\newcommand{\sg}{\sigma}
\newcommand{\rg}{\rho}

\newcommand{\fg}{\phi}
\newcommand{\vf}{\varphi}
\newcommand{\og}{\omega}

\newcommand{\Del}{\Delta}

\newcommand{\Og}{\Omega}

\newcommand{\lh}{\left(}
\newcommand{\rh}{\right)}

\begin{document}

\pagestyle{empty} 
\begin{flushright} 
NIKHEF/02-004
\end{flushright}
\vs{3}
\begin{center} 
{\Large{\bf Cosmic scalar fields with flat potential$^{*}$}} \\
\vs{7} 

{\large J.W.\ van Holten }\\
\vs{2} 

{\large NIKHEF, Amsterdam NL$^{**}$}\\ 
\vs{3} 

June 23, 2002
\vs{15} 

{\small{\bf Abstract}}
\end{center} 

\nit
{\footnotesize{The dynamics of cosmic scalar fields with flat potential is studied.  
Their contribution to the expansion rate of the universe is analyzed, and their 
behaviour in a simple model of phase transitions is discussed. }} 
\vfill
\footnoterule 
\nit
{\footnotesize{
$^*$Work supported by the research program FP52 of the Foundation for 
Fundamental Research of Matter (FOM). \\
$^{**}$e-mail: \tt{ v.holten@nikhef.nl}}}
\np 
~\hfill

\np

\pagestyle{plain}
\pagenumbering{arabic} 

\nit
The universe is observed to expand \ct{hubble}, and this expansion is suspected to have 
accelerated at a late stage of its evolution \ct{perlm,hkz}. Furthermore,
observation of the cosmic microwave background indicates the universe is spatially flat
\ct{comb}. The energy density of scalar fields has been recognized to contribute to the 
expansion of the universe \ct{linde,dreitl,veltman2}, and has been proposed 
to explain inflation \ct{guth}, as well as the presently observed expansion 
\ct{wetterich,peebles,steinhardt1,stmukh,wett2}; in  ref.\ \ct{starobinsky} an attempt has 
been made to confront the data with the predictions for a minimally coupled scalar field 
with an a priori unknown potential. An overview of models with many references has 
been given in \ct{sahni}.

As suggested by the standard model of the electroweak interactions and its many proposed
small-distance extensions with or without supersymmetry, the dynamics of the universe 
involves a large number of scalar fields interacting with gravity, gauge fields and 
fermionic matter. The observed isotropy and homogeneity of the universe do not 
allow for the existence of long-range electric and magnetic fields, but neutral 
scalar fields can have non-trivial dynamics in an expanding FRW-type universe.  

For a number of minimally coupled scalar fields $\fg_i$ with a potential $V[\fg]$ 
the relevant dynamical equations for a flat universe $(k = 0)$ are 
\be 
\ddot{\fg}_i + 3 H \dot{\fg}_i + V_{,i} = 0, \hs{2} 
\frac{1}{2}\, \sum_i \dot{\fg}_i^2 + V[\fg] + \rg = \frac{3H^2}{8\pi G}, 
\label{1.1}
\ee 
where $H$ is the Hubble parameter, and $\rg$ is the contribution of matter to the 
density of the universe. It should be noted, that we absorb any cosmological constant in 
a constant contribution to the potential $V[\fg]$. 
\vs{1}

\nit
{\em Constant scalar fields.} 
As observed in \ct{linde,dreitl,veltman2} simple non-trivial solutions to the coupled equations 
(\ref{1.1}) are provided by constant fields at stationary points of the potential:
\be 
V_{,i} = 0, \hs{2} V_m + \rg = \frac{3H^2}{8\pi G}. 
\label{1.2}
\ee 
Here $V_m$ is the value of the potential at the stationary point (presumably a minimum), 
and $\rg$ represents the density of ordinary relativistic and cold non-relativistic matter, with 
present densities $\rg_{r, 0}$ and $\rg_{nr, 0}$. Then  the last 
equation becomes in terms of the scale factor $\sg(t) = a(t)/ a_0$: 
\be 
\dot{\sg}^2 = h^2 \sg^2 + \frac{8\pi G}{3}\, \lh \frac{\rg_{r,0}}{\sg^2}\, + 
 \frac{\rg_{nr,0}}{\sg} \rh,  \hs{2} 
 h^2 = \frac{8\pi G V_m}{3}, 
\label{1.3}
\ee 
provided $V_m > 0$. Hence for large universes the late-time behaviour of the scale factor is 
described by 
\be 
a(t) \sim a_0 e^{h(t - t_0)}.
\label{1.4}
\ee 
This is consistent with the observed present state of the universe if the Hubble parameter is
$h \simeq 70$ km/sec/Mpc, of the order of the inverse life time of the universe. As is well-known, 
the corresponding value of $V_m$ is extremely small: $V_m = \rg_c \simeq 10^{-122}$ 
in Planck units. 

In contrast, for $V_m = 0$, we obtain the standard radiation/matter dominated solutions with 
$a(t) \sim t^{\kg}$, where $\kg = 1/2$ or $\kg = 2/3$. If $V_m = - 3\, \og^2/(8\pi G) 
<0$, there are oscillating solutions with maximal amplitude $\bar{a} =  \bar{\sg} a_0$ given 
by the solution of 
\be 
\og^2 \bar{\sg}^4 - \frac{8\pi G}{3} \lh \rg_{nr,0}\, \bar{\sg} + \rg_{r,0} \rh = 0.
\label{1.5}
\ee 
The last two cases can be consistent with a flat universe only if the density of matter 
and radiation equals or exceeds the critical density $\rg_c$, which is definitely contradicted
by the best current estimates of the densities of luminous and cold dark matter. 
\vs{1} 

\nit
{\em Flat potentials.} 
A different evolution of the Hubble parameter $H(t)$ is obtained by allowing one or 
more scalar fields to depend on time. We consider the situation where the first field is 
dynamical: $\fg_1 = \vf(t)$, whilst all other fields are constant: $\fg_i = v_i$ 
$(i = 2,3,...,N)$. Then $V(\vf) = V[\fg_1 = \vf; \fg_i = v_i]$. Thus we have 
\be 
\ddot{\vf} + 3H \dot{\vf} + V^{\prime} = 0, \hs{2} 
 \frac{1}{2}\, \dot{\vf}^2 + V(\vf) + \rg = \frac{3H^2}{8\pi G}. 
\label{1.6}
\ee 
In general any solution depends sensitively on the initial value $\vf_{init}$, which 
is difficult to control. However, this problem is absent if the potential is invariant
under field translations $\tilde{\vf} = \vf + \eps$. Therefore it is of some interest 
to study flat potentials $V(\vf) = V_0$. Such potentials are characteristic for Goldstone 
bosons, and they occur in many supersymmetric models. From the flatness of $V$
one derives a conservation law for the field momentum per co-moving volume:
\be 
\ddot{\vf} + 3 H \dot{\vf} = 0 \hs{1} \Rightarrow \hs{1} 
\sg^3 \dot{\vf} = \gam = \mbox{constant}.
\label{1.7}
\ee 
After multiplication by $\sg^6$ the other equation then becomes 
\be 
\frac{\gam^2}{2}\, + \rg \sg^6 + V_0  \sg^6 = \frac{3}{8\pi G}\, \lh \sg ^3 H\rh^2
 = \frac{1}{24\pi G}\, \lh \frac{d\sg^3}{dt} \rh^2. 
\label{1.8}
\ee 
The solutions can be characterized qualitatively as in the case of constant scalar fields: \nl 
a.\ For $V_0 > 0$, the potential provides the leading term, and for large $a$ we 
find as before:
\be 
a(t) \simeq a_0\, e^{h (t - t_0)}, \hs{2} h = \sqrt{\frac{8\pi GV_0}{3}}. 
\label{1.9}
\ee 
b.\ For $V_0 = 0$, the leading term is provided by the matter density $\rg$, and 
the evolution of the scale factor is again of the form $a(t) \sim t^{\kg}$, where 
$\kg = 1/2$ for relativistic matter, and $\kg = 2/3$ for cold non-relativistic matter. \nl
c.\ Finally, for $V_0 < 0$ there are oscillating solutions, with the maximal 
scale $\bar{a} = \bar{\sg} a_0 $ a solution of
\be 
\frac{\gam^2}{2}\, + \bar{\rg}\, \bar{\sg}^6 + V_0\, \bar{\sg}^6 = 0,
\label{1.10}
\ee 
which is a direct generalization of eq.(\ref{1.5}). 
\vs{1} 

\nit
{\em Matter-dominated regime.} 
The precise evolution of the scale factor and the scalar field can be solved analytically 
in the matter-dominated regime, in which $\rg a^3 =$ constant. In the following we 
describe the solutions in some detail. 

It is convenient to use the dimensionless scale variable $\sg = a/a_0 = 1/(1 + z)$. 
Assuming $V_0 \neq 0$, eq.(\ref{1.8}) then becomes
\be 
\lh \frac{d\sg^3}{dt} \rh^2 = 24 \pi G V_0 \left[ \lh \sg^3 + \frac{\rg_0}{2V_0} \rh^2 + 
 \frac{\gam^2}{2V_0} - \frac{\rg_0^2}{4V_0^2} \right], 
\label{2.1}
\ee 
where  $\rg(\sg) = \rg_0/\sg^3$. As this equation is invariant under 
time reversal, at any time we can have an increasing and a decreasing solution; we always exhibit
here the solution with growing scale factor at earely times. For positive values of $V_0$ we must 
distinguish three cases: \nl 
a.  $\gam^2 > \rg_0^2/2 V_0 > 0$. Then the solution of eq.(\ref{2.1}) is 
\be 
\sg(t) = \frac{a(t)}{a_0} = \left[  A\, \sinh 3 h  (t - t_0 - \Del)  - \frac{\rg_0}{2V_0} \right]^{1/3}, 
\label{2.2}
\ee 
with 
\be 
A = \sqrt{\frac{\gam^2}{2V_0} - \frac{\rg_0^2}{4V_0^2}}, \hs{2} 
3h\Del = - \mbox{arcsinh} \left[ \frac{1}{A}\, \lh 1 + \frac{\rg_0}{2V_0} \rh \right], 
\label{2.2.0}
\ee  
and $h$ given in eq.(\ref{1.9}). The corresponding solution for the scalar field is 
\be 
\ba{rcl}
\vf(\sg) & = & \dsp{ \vf_0 - \frac{1}{\sqrt{12\pi G}}\, \lh \mbox{arcsinh} \left[ 
 \frac{1}{\sg^3}\, \frac{\gam^2 + \rg_0\, \sg^3}{\sqrt{2\gam^2 V_0 - \rg_0^2}}\,
 \right] - \Og \rh, }\\ 
\ea 
\label{2.2.1}
\ee 
with the intergration constant given by 
\be
\Og  = \mbox{arcsinh} \left[ \frac{\gam^2 + \rg_0}{\sqrt{2\gam^2 V_0 - \rg_0^2}}\, \right].
\label{2.2.2}
\ee 
As all equations are invariant under reflections of the scalar field: $\vf  \rightarrow -\vf$,
there is another solution
\be
\vf(\sg) = \vf_0 + \frac{1}{\sqrt{12\pi G}}\, \lh \mbox{arcsinh} \left[ 
 \frac{1}{\sg^3}\, \frac{\gam^2 + \rg_0\, \sg^3}{\sqrt{2\gam^2 V_0 - \rg_0^2}}\,
 \right] - \Og \rh. 
\label{2.2.3}
\ee 
b. $\gam^2 = \rg_0^2/2 V_0$. In this case the constant terms in the square brackets in (\ref{2.1}) 
cancel, with the result that
\be 
\sg(t) = \left[ 1 + \lh 1 + \frac{\rg_0}{2V_0} \rh \lh e^{3 h (t - t_0)} - 1 \rh  
  \right]^{1/3}.
\label{2.3}
\ee 
The solution for the scalar field reads  
\be
\vf(\sg) = \vf_0 \mp \frac{1}{\sqrt{12\pi G}}\, 
 \ln \left[ \frac{1}{\sg^3}\, \frac{\gam^2 + \rg_0\, \sg^3}{\gam^2 +\rg_0} \right]. 
\label{2.3.1}
\ee 
To obtain the coefficients in front of and inside the logarithm, we have used the relation between 
$V_0$ and $h$, and the special relation between $V_0$ and $\rg_0$ and $\gam$. \nl
c. $ \gam^2 < \rg_0^2/2 V_0$. The solution of eq.(\ref{2.1}) becomes 
\be 
\sg(t) = \left[ A\, \cosh [3 h (t - t_0- \Del)]  -  \frac{\rg_0}{2V_0} \right]^{1/3},
\label{2.4}
\ee 
with 
\be
A = \sqrt{\frac{\rg_0^2}{4V_0^2} - \frac{\gam^2}{2V_0}}, \hs{2}   
3h\Del = \mbox{arccosh} \left[ \frac{1}{A}\, \lh 1 + \frac{\rg_0}{2V_0} \rh \right].
\label{2.4.0}
\ee 
The scalar field evolution is given by   
\be
\vf(\sg) = \vf_0  \mp \frac{1}{\sqrt{12 \pi G}}\, \lh \mbox{arccosh}\, \left[ \frac{1}{\sg^3}\, 
 \frac{\gam^2 + \rg_0\, \sg^3}{\sqrt{\rg_0^2 - 2V_0 \gam^2}} \right] - \Og \rh, 
\label{2.4.1}
\ee 
where the constant $\Og$ is 
\be 
\Og = \mbox{arccosh}\, \left[ \frac{\gam^2 + \rg_0}{\sqrt{\rg_0^2 - 2V_0 \gam^2}} \right].
\label{2.4.2}
\ee 
Observe, that in all three solutions above $\vf$ diverges for $\sg \rightarrow 0$, as does  
$\dot{\vf}$. However, long before this limit is reached the universe becomes relativistic, and 
our model must be modified. The initial conditions for the model discussed here should 
rather be set at the onset of the matter-dominated aera. 
\nl
d. If $V_0 = 0$ equation (\ref{1.8}) implies 
\be 
\sg(t) = \left[ 1 +\sqrt{12\pi G \lh 2\rg_0  + \gam^2 \rh}\, 
 \lh t - t_0 \rh + 6\pi G \rg_0 \lh t - t_0 \rh^2 \right]^{1/3}. 
\label{2.5}
\ee 
(We have chosen the solution which can vanish at $t = 0$.) This is the usual cold matter 
dominated universe. The scalar field equation is most easily solved directly in terms of time: 
\be 
\vf(t) =  \vf_0  \mp \frac{1}{\sqrt{3\pi G}}\, \lh \mbox{arctanh} 
 \left[ \frac{\sqrt{12\pi G}\, \rg_0 (t - t_0 + \Del)}{\gam}\, \right] -\Og \rh, 
\label{2.5.1}
\ee 
with 
\be  
\Del =  \sqrt{\frac{\gam^2 + 2\rg_0}{12\pi G \rg_0^2}}, \hs{2} 
\Og = \mbox{arctanh}\, \sqrt{1 + \frac{2\rg_0}{\gam^2}}. 
\label{2.5.0}
\ee 
This solution is asymptotically constant; 
\be
\vf_{\infty} = \vf_0 \pm \frac{1}{\sqrt{3\pi G}}\, \lh \mbox{arctanh} 
 \left[ \sqrt{1 + \frac{2\rg_0}{\gam^2}}\, \right]  - 1\rh. 
\label{2.5.2}
\ee  
The field can have reached its asymptotic value at present only if  $\gam^2 \ll \rg_0$. \nl  
e.  For $V_0 < 0$,  the scale factor takes the periodic form 
\be 
\sg(t) = \left[ \frac{\rg_0}{2|V_0|}\, + A\, \sin 3 \og (t - t_0 - \Del) \right]^{1/3}, 
\label{2.6}
\ee 
with 
\be 
A = \sqrt{\frac{\rg_0^2}{4V_0^2} + \frac{\gam^2}{2|V_0|}}, 
\hs{2} \og = \sqrt{\frac{8\pi G |V_0|}{3}}, 
\label{2.6.1}
\ee 
and 
\be 
3 \og \Del = \mbox{arcsin} \left[ \frac{1}{A} \lh \frac{\rg_0}{2|V_0|} - 1 \rh \right]. 
\label{2.6.0}
\ee 
The scalar field solution reads
\be 
\vf(\sg) = \vf_0  \mp \frac{1}{\sqrt{12\pi G}}\, \lh\mbox{arccosh} \left[ \frac{1}{\sg^3}\, 
 \frac{\gam^2 + \rg_0 \sg^3}{\sqrt{2|V_0| \gam^2 + \rg_0^2}} \right] - \Og \rh, 
\label{2.7}
\ee 
where 
\be 
\Og = \mbox{arccosh} \left[ \frac{\gam^2 + \rg_0 }{\sqrt{2|V_0| \gam^2 + \rg_0^2}} \right].
\label{2.8}
\ee 
Actually, this is formally the same solution as for the scalar field in case (c). However, it 
evolves in a universe with a maximum expansion and a finite life time, rather than an 
infinitely expanding one. 
\vs{1} 

\nit
{\em Equation of state}. 
As cold matter is pressureless, the total pressure of the cosmic fluid comes from the scalar 
field:
\be 
p_{tot} = p_s = \frac{\gam^2}{2\sg^6} - V_0. 
\label{3.1}
\ee 
The total density is 
\be 
\rg_{tot} = \rg_s + \rg = \frac{\gam^2}{2\sg^6} + \frac{\rg_0}{\sg^3} + V_0. 
\label{3.2}
\ee 
It follows that the state parameter is given by 
\be
w_{tot} = \frac{p_{tot}}{\rg_{tot}}\, 
 = - \frac{2 V_0 \sg^6 - \gam^2}{2V_0 \sg^6 + 2 \rg_0 \sg^3 + \gam^2}. 
\label{3.3}
\ee 
For infinitely expanding universes with $V_0 > 0$, in the large-scale limit the parameter 
becomes $w_{\infty} = -1$. Inflationary behaviour sets in when $w_{tot} < - 1/3$. 
This happens at a critical scale $\sg_e$ given by 
\be 
\sg_e^3 = \frac{\rg_0}{4V_0}\, \lh 1 + \sqrt{1 + \frac{16\gam^2 V_0}{\rg_0^2}} \rh. 
\label{3.4}
\ee 
Depending on the relative values of $V_0$ and $\rg_0^2/2\gam^2$, we have 
\be 
\ba{rcl} 
\gam^2 > \rg_0^2/2 V_0 & \rightarrow & \sg_e^3 > \frac{\rg_0}{V_0}; \\ 
 & & \\
\gam^2 = \rg_0^2/2 V_0 & \rightarrow & \sg_e^3 = \frac{\rg_0}{V_0}; \\ 
 & & \\
\gam^2 < \rg_0^2/2 V_0 & \rightarrow & \sg_e^3 < \frac{\rg_0}{V_0}. \\ 
\ea 
\label{3.5}
\ee 
For vanishing potential $V_0 = 0$ the parameter is always positive, but approaches 
$w_{tot} \downarrow 0$ in the large-scale limit. For $V_0 < 0$ the state parameter 
$w_{tot}$ is always positive and becomes infinite at the turning point $\sg = 
\bar{\sg}$. 

The ratio of dark energy to cold matter is given by 
\be 
y = \frac{\rg_s}{\rg}\, = \frac{\gam^2}{2\rg_0}\, \frac{1}{\sg^3} + \frac{V_0}{\rg_0}\, \sg^3.
\label{3.6}
\ee 
For $V_0 > 0$ it never becomes small, but it reaches a minimum for $\sg_m^6 = \gam^2/2V_0$,
when 
\be 
y_m = \sqrt{\frac{2V_0 \gam^2}{\rg_0^2}}. 
\label{3.7}
\ee 
Thus a situation in which $y_m < 1$ can be reached only if $\gam^2 < \rg_0^2/2V_0$. In 
this case
\be 
\frac{\sg_e^3}{\sg_m^3} < \frac{2}{y_m}.
\label{3.8}
\ee 
From this analysis we observe, that a recent acceleration of the universe conjectured on the basis of 
the type Ia supernova data can be incorporated in these simple models for $V_0 > 0$. By 
construction the development of the scalar field does not depend strongly on the initial conditions,   
as in more sophisticated models with tracking behaviour \ct{steinhardt2}. Also, we can arrange 
for the scalar density to be comparatively low in the early universe, and have accelerated expansion 
at a later time if we take the parameters in the domain $0< \gam^2 < \rg_0^2/2 V_0$.  
\vs{1} 

\nit
{\em Piece-wise constant potentials.} The analysis of flat potentials can be extended in a 
straightforward way to the case of piece-wise constant potentials. The main new feature 
is that whenever the field reaches a potential step, we must determine and match its behaviour 
before and after it hits the discontinuity. This is done by rewriting the field equation in the
form
\be
\frac{d}{dt} \lh \sg^3 \dot{\vf} \rh = - \sg^3 V^{\prime} \hs{1} \Rightarrow \hs{1} 
 \frac{d}{dt} \lh \frac{1}{2} (\sg^3 \dot{\vf})^2 \rh = - \sg^6 \frac{dV}{dt}.
\label{6.0}
\ee 
The second equation can be integrated under two different conditions: \nl
(i) if the field crosses the potential step ({\em transmission}), we obtain 
\be 
\frac{\Del \gam^2}{2\bar{\sg}^6} = - \Del V,
\label{6.0.1}
\ee
with $\Del V = V_2 - V_1$ the magnitude of the potential step, $\bar{\sg}$ the scale
at which the field reaches the discontinuity in the potential, and $\Del \gam^2 = \gam_2^2
- \gam_1^2$ the change in the square of the  field momentum; \nl
(ii) if the field does not cross the potential step ({\em reflection}), the value of the potential 
does not change, and $\gam^2$ is equal before and after the reflection. As the 
field now must run in the other direction as a function of time, we deduce that
$\gam_2 = - \gam_1$. 

The condition (\ref{6.0.1}) can also be written in the form 
\be 
\frac{\gam_1^2}{2\bar{\sg}^6} + V_1 = \frac{\gam_2^2}{2\bar{\sg}^6} + V_2.
\label{6.0.2}
\ee 
As matter is not created nor destroyed, this shows, that  $H$ is continuous during the change in 
the scalar field. In contrast, during transmission across the potential step the derivative 
of the field $\dot{\vf}$, the derivative of the Hubble parameter $\dot{H}$ and the potential all  
change by a finite amount.  This can also be derived by taking a derivative of the second eq.(\ref{1.6}),  
which leads to the equations 
\be 
-\frac{\dot{H}}{8\pi G} = \frac{1}{2}\, \dot{\vf}^2 + \frac{1}{2}\, \rg, \hs{2}
 \frac{3H^2 + \dot{H}}{8\pi G}\, = V + \frac{1}{2}\, \rg. 
\label{6.1}
\ee 
From the continuity of $\rg$ and $H$, and the finite discontinuity of $V$, the finite
discontinuous changes in $\dot{H}$ and $\dot{\vf}$ are obtained directly:
\be
\Del V = \frac{\Del \dot{H}}{8\pi G} = -\Del \lh \frac{1}{2} \dot{\vf}^2\rh
 = - \frac{\Del(\gam^2)}{2\bar{\sg}^6}.
\label{6.2}
\ee 
The same equations actually also hold trivially at reflection, as all changes in these
quantities then vanish. In this case continuity of the field is obtained by taking a solution $\vf(\sg) = 
\vf_1(\sg)$ before the reflection ($\sg < \bar{\sg}$), and $\vf(\sg) = \vf_2(\sg)$ after 
the reflection ($\sg > \bar{\sg}$), with the two solutions related by 
\be 
\vf_2(\sg)= 2 \bar{\vf} - \vf_1(\sg), \hs{2} \bar{\vf} = \vf_1(\bar{\sg}) = \vf_2(\bar{\sg}). 
\label{6.2.1}
\ee 
Observe, that if the initial condition is such that 
\be
\frac{\gam_1^2}{2\bar{\sg}^6} + V_1 < V_2,
\label{6.5}
\ee 
then only the reflected solution exists.  

Obviously, the piece-wise constant potentials discussed here can be used as approximations 
for a variety of smoothly changing potentials. It is to be noted that the translation symmetry
in field space is then lost, but one  still has relative freedom in choosing initial conditions 
in any flat region of the potential. 
\nl

\nit
{\em Phase transitions.} 
In the course of evolution of the universe, phase transitions accompanied by symmetry 
breaking can change the value of the potential. We consider the simple situation that
the potential changes from a value $V_1$ before the transition time $t_c$ to a value 
$V_2$ after $t_c$. For simplicity we take the transition to occur instantaneously. 
We can then treat the phase transition as a potential step occurring at the time $t_c$.
Eq.(\ref{6.0.2}) then implies, that during the transition 
\be 
\frac{\gam_1^2}{2\sg_c^6} + V_1 = \frac{\gam_2^2}{2\sg_c^6} + V_2, 
\label{4.3}
\ee 
where $\sg_c = \sg(t_c)$. Note, that in this case we only have the possibility of transmission, 
not reflection of the field. As the matter density and Hubble parameter are not affected during the 
transition, the continuity of $\sg(t)$ and $\vf(t)$ allow us to express the parameters of the solution 
for $t > t_c$ in terms of those of the solution for $t < t_c$.   

In this context, consider again the parameter $y_m = \sqrt{2\gam^2 V/\rg_0^2}$, 
the minimal ratio of scalar to matter density. From eq.(\ref{4.3}) it follows, that 
after a phase transition
\be 
y^2_{m2} - y^2_{m1} = \frac{2\sg_c^6}{\rg_0^2}\, \lh V^2_1 - V^2_2 \rh
 + \frac{1}{2\rg_0^2 \sg_c^6}\, \lh \gam_1^4 - \gam_2^4 \rh.
\label{4.4}
\ee 
Thus, whether the transition leads to a situation with larger or smaller $y_m$ depends 
both on the difference in potential energy density as well as on the difference in
kinetic energy density of the field. In particular, the value remains the same if
\be 
\frac{V_1}{V_2} = \frac{\gam_2^2}{\gam_1^2} \hs{1} \Rightarrow \hs{1} 
 \frac{\gam_1^2}{2\sg_c^6} = V_2, \hs{1} \frac{\gam_2^2}{2\sg_c^6} = V_1,
\label{4.5}
\ee 
i.e.\ if the values of kinetic and potential energy are interchanged during the transition. 
In all cases where this condition is not met, the value of $y_m$ changes. If the potential
energy density dominates over the kinetic energy density, than lowering $V$ will in general 
lower $y_m$, whereas in kinetic-energy dominated situations a lowering of $V$ will give 
rise to a larger value of $y_m$ after the transition. This follows from the observation, that
eq.(\ref{4.4}) can be replaced by the  equivalent expression
\be 
y_{m2}^2 - y_{m1}^2 = \frac{2\sg_c^6}{\rg_0^2}\, \lh V_1 - V_2 \rh 
 \lh V_1 + V_2 - \frac{\gam_1^2 + \gam_2^2}{2\sg_c^6} \rh. 
\label{4.6}
\ee 
As generically the nature of the solution for
$\sg(t)$ and $\vf(t)$ depends on whether $y_m^2$ is positive, zero or negative, and
in the first case on whether it is equal to,  larger or less than unity, the 
solutions before and after a phase transition may belong to different classes a-e discussed above. 
It should be noted that the analysis also applies to the case where an originally constant field 
($\gam_1 = 0$) becomes dynamical ($\gam_2 \neq 0$), in which case 
$V_1 = V_2 +\gam_2^2/2\sg_c^6$. 

In conclusion, massless scalar fields associated with flat potentials display simple but 
interesting cosmological dynamics. As they are associated with Goldstone bosons, 
they can play a role in phase transitions during which they change their dynamics. 
Finally, it has been shown how to piece together the solutions for potentials with 
discrete jumps, which under certain conditions can give qualitative insight in the behaviour 
of  non-constant  potentials.

\end{document}